\documentclass[11pt,a4paper]{article}
\usepackage{jheppub}

\usepackage{amsmath}
\usepackage{amsfonts}
\usepackage{amssymb}
\usepackage{dsfont}
\usepackage{accents}
\usepackage[mathscr]{eucal}
\usepackage{braket}

\usepackage{enumitem}

\usepackage{epsfig}
\usepackage{graphicx}

\newcommand{\rmd}{\ensuremath{\mathrm{d}}}
\newcommand{\rme}{\ensuremath{\mathrm{e}}}
\newcommand{\rmi}{\ensuremath{\mathrm{i}}}

\newcommand{\vb}{\ensuremath{{\bar{v}}}}
\newcommand{\ub}{\ensuremath{{\bar{u}}}}

\newcommand{\rs}{\ensuremath{r_{\rm s}}}

\newcommand{\ks}{\ensuremath{\kappa^{\rm{st}}}}
\newcommand{\vl}{\ensuremath{v_l}}
\newcommand{\apr}{\ensuremath{a_{\rm{p}}}}

\title{Hawking versus Unruh effects, or the difficulty of slowly crossing a black hole horizon}

\author[a,b]{Luis C.\ Barbado,}
\author[b]{Carlos Barcel\'o,}
\author[c,d]{Luis J.\ Garay}
\author[e]{and Gil Jannes}

\affiliation[a]{Quantenoptik, Quantennanophysik und Quanteninformation, Fakult\"at f\"ur Physik, Universit\"at Wien, Boltzmanngasse 5, 1090 Wien, Austria}
\affiliation[b]{Instituto de Astrof\'isica de Andaluc\'ia (CSIC), Glorieta de la Astronom\'ia, 18008 Granada, Spain}
\affiliation[c]{Departamento de F\'isica Te\'orica II, Universidad Complutense de Madrid, 28040 Madrid, Spain}
\affiliation[d]{Instituto de Estructura de la Materia (CSIC), Serrano 121, 28006 Madrid, Spain}
\affiliation[e]{Departamento de Ciencias y Tecnolog\'ia, Universidad Europea de Madrid, C/ Tajo, 28670 Villaviciosa de Od\'on, Madrid, Spain}

\emailAdd{luis.cortes.barbado@univie.ac.at}
\emailAdd{carlos@iaa.es}
\emailAdd{luisj.garay@ucm.es}
\emailAdd{gil.jannes@universidadeuropea.es}

\abstract{When analyzing the perception of Hawking radiation by different observers, the Hawking effect becomes mixed with the Unruh effect. The separation of both effects is not always clear in the literature. Here we propose an inconsistency-free interpretation of what constitutes a Hawking effect and what an Unruh effect. An appropriate interpretation is important in order to elucidate what sort of effects a detector might experience depending on its trajectory and the state of the quantum field. Under simplifying assumptions we introduce an analytic formula that separates these two effects. Armed with the previous interpretation we argue that for a free-falling detector to cross the horizon without experiencing high-energy effects, it is necessary that the horizon crossing is not attempted at low velocities.}

\keywords{Black holes, Hawking radiation, Unruh effect, Quantum Field Theory in Curved Spacetime, Vacuum states}

\arxivnumber{1608.02532}

\begin{document}

\maketitle

\section{Introduction}

Arguably, the two cornerstone results of Quantum Field Theory in curved spacetimes and non-inertial reference frames are the Hawking effect in black holes~\cite{Hawking:1974sw} and the Unruh effect~\cite{Unruh:1976db}. The Unruh effect is typically considered a \emph{subjective} effect, meaning that it is something that is not objectively there but shows up as detector perception effects when subject to acceleration; it is typically presented, in its most basic form, as the perception of an accelerated detector in Minkowski spacetime. On the other hand, the Hawking effect is interpreted as an \emph{objective} effect, something that happens to any black hole once it is formed. 
For an observer following an arbitrary trajectory outside a black hole, it is clear that these two effects will in general be present together; one could then talk about perception effects near radiating black holes (see e.g.~\cite{Barbado:2011dx,frolov}). 

When looking at the literature on radiating black holes, however, there is often no consensus nor clarity on what constitutes a Hawking effect and what constitutes an Unruh effect in the net perception of the observer (see e.g.\ the discussions in~\cite{Singleton:2011vh,Crispino:2012zz, Smerlak:2013sga,Singh:2014paa,Greenwood:2008zg,Eune:2014eka,Kim:2016iyf}; even the very existence of the Unruh effect is sometimes still questioned, see e.g.~\cite{Gelfer:2015voa}). A controversy on whether the equivalence principle is preserved in the presence of these effects is still unsettled~\cite{Singleton2016}. Our work can help to clear up this controversy. More generally, the existence or not of a separation between the Hawking and Unruh effects, and why and how this separation should be accomplished is the main theme of the present work.

To see the problem clearly, let us consider an observer sustaining himself at rest at a fixed radius just outside the horizon of a radiating black hole. A standard calculation concludes that this observer perceives an outgoing flux of black-body radiation at the Hawking temperature associated with the black hole, corrected by an enormous gravitational blueshift due to the location of the observer. It is commonly  stated that this tremendous temperature can be interpreted as an Unruh effect (this association can be read of e.g.\ in Birrell-Davies book, page 282~\cite{Birrell:1982ix}). As we will argue, this association is potentially misleading. The standard argument is that the Unruh vacuum state (the state of a radiating black hole) is perceived as vacuum by free-falling observers at the horizon; then, as the observer is strongly accelerating with respect to the local free-falling reference frame in order to sustain his position, he must experience an Unruh effect and thereby perceive a thermal bath with a temperature proportional to his acceleration. This reasoning indeed yields the correct result for the temperature (strictly speaking, only in the horizon limit; see the discussion in~\cite{Barbado:2011dx}). The problems with this interpretation come about when one asks oneself whether this radiation could or could not lead to some buoyancy effect, that is, when one tries to understand the feedback effects of the radiation on the detector trajectory. Interpreted as an Unruh effect, one will immediately conclude that the answer is negative: Any back-reaction associated with the Unruh effect should work against maintaining the level of acceleration and thus add to the gravitational pull of the black hole. On the other hand, interpreted as a Hawking effect---that is, a radiation objectively produced by the black hole and augmented by the gravitational blueshift due to the position---,  the radiation coming from the horizon and impinging on the detector will lead to radiation pressure and could thus indeed produce some
buoyancy effect.

Our position is that the latter case is what would actually occur, as
first discussed by Unruh and Wald~\cite{UnruhWald1982} (see also
Bekenstein~\cite{Bekenstein1999}). Thus, we argue that the first
interpretation is physically inappropriate. We must say that for instance
Wald~\cite{Wald1994,Wald1999} has pointed out since many years that the Unruh and
Hawking effects are quite distinct. Indeed, in the Hawking effect there is only an outgoing flux, while in the Unruh effect the flux is bidirectional, leading in practice to a bath rather than to a flux. Then, he puts in parallel
an Unruh effect with a black hole in the Hartle-Hawking state, as opposed
to one in the Unruh state. However, the previous interpretational problem
remains. Bekenstein used a conceptualization based on the Hartle-Hawking
state to calculate buoyancy forces due to gradients of
pressure~\cite{Bekenstein1999}. Again, if one were to understand this as a pure Unruh effect,
the tendency of the effect should be quite the opposite, namely a reinforcement of the pull of gravity.

In this work we propose a different interpretation of the Unruh effect and a
way of separating the Hawking and Unruh effects that is free of
insconsistencies. The separation is clear-cut under the simplifying assumptions of our work: presence of an asymptotically flat region and negligible back-scattering effects. Although in more general situations the separation will not be so clear-cut we think that our interpretation will still serve as a guiding principle. Our approach to this problem is based on the \emph{effective temperature function,} a function introduced in~\cite{Barcelo:2010pj,Barcelo:2010xk} which accounts, under certain adiabaticity conditions, for the temperature of the radiation perceived by an observer in a given vacuum state. 

The advocated interpretation will also allow us to revisit and reinforce some results that we previously obtained in~\cite{Barbado:2011dx, Barbado:2012pt, Barbado:2015zga, Barbado2016}, as well as to obtain several new results. Of special interest, we will discuss that a combined effect of Hawking and Unruh radiation over the trajectory of a test object, could make it difficult for the object to cross the horizon when this crossing is attempted at low velocities. It appears that for a free-falling observer to experience nothing peculiar when crossing a black hole horizon, as is commonly assumed, a large horizon-crossing velocity is required. Interestingly, this result resonates with other results in black hole physics in which the horizon has to be formed with sufficient swiftness for quantum effects not to disturb the standard semiclassical picture (see the discussion in~\cite{Barcelo2008}). To avoid any confusion, let us note that these effects are completely unrelated to the possible existence of a physical firewall~\cite{Almheiri:2012rt}.

The structure of this paper is the following. We will set the stage in Sec.~\ref{sec:prelim}, with some general preliminaries. In Sec.~\ref{sec:H-vs-U} we discuss the separation between Hawking radiation and the Unruh effect. Sec.~\ref{sec:examples} contains the calculation of the radiation perception for two important categories of observers: static observers and radially free-falling observers in Schwarzschild spacetime. The calculations for these observers will allow us to deduce an important consequence of this work: the difficulty of crossing the event horizon of a black hole at low velocities, which we discuss in Sec.~\ref{sec:difficult}. Finally, we summarize and discuss our results in Sec.~\ref{sec:summary}.

\section{Preliminaries}\label{sec:prelim}

We will focus the discussion on the most fundamental and irreducible characteristics of Hawking radiation and the Unruh effect. Thus, we work with the simplest possible scenario: a conformally invariant massless real scalar field in a $(1+1)$-dimensional spacetime. This amounts to neglect any backscattering by the geometry. In addition we will assume the existence of an asymptotically flat region. The scalar field satisfies the massless Klein-Gordon equation, which in some fiduciary null coordinates $(\ub,\vb)$ reads
\begin{equation}
\square \phi=	\frac{\partial}{\partial \ub} \frac{\partial}{\partial \vb} \phi = 0.
\label{klein-gordon}
\end{equation}
Conformal invariance makes the~$\ub$ and~$\vb$ radiation sectors decouple, with the general solution being $\phi (\ub,\vb)= f(\ub) + g(\vb)$. It also allows any relabeling $U=U(\ub )$, \mbox{$V=V(\vb)$} to be associated with a corresponding decomposition of the field in terms of orthogonal modes $\propto \{\rme^{-\rmi \omega  U}, \rme^{-\rmi \omega  V}\}$, $\omega > 0$. After a canonical quantization procedure, we obtain natural annihilation $\hat a_{\omega }^U,\hat a_{\omega }^V$ and creation ${\hat a_{\omega }^U}^\dag,{\hat a_{\omega}^V}^\dag$ operators and a natural vacuum state~$\ket{0}$ associated with these modes.

\subsection{Effective temperature function}

In order to know the perception of a generic observer in the vacuum state~$\ket{0}$, one can compute the Bogoliubov transformation between the modes defining the vacuum state and the modes to which the observer naturally couples. The effective temperature function is based on this Bogoliubov transformation~\cite{Barcelo:2010pj,Barcelo:2010xk}. Given an observer with proper time~$\tau$ following a trajectory~$(U(\tau),V(\tau))$, the effective temperature functions for the~$U$ and~$V$ radiation sectors are, respectively,
\begin{equation}
\kappa_U (\tau):= -\left. \frac{\rmd^2 U}{\rmd \tau^2} \middle/ \frac{\rmd U}{\rmd \tau} \right., \qquad  \kappa_V (\tau):= -\left. \frac{\rmd^2 V}{\rmd \tau^2} \middle/ \frac{\rmd V}{\rmd \tau} \right.. 
\label{kappa}
\end{equation}
When one of these functions remains approximately constant for a sufficiently long period of time, the observer perceives a planckian spectrum of particles during this period (in the corresponding radiation sector) with temperature $T = |\kappa_U| / (2 \pi)$ and $T = |\kappa_V| / (2 \pi)$, respectively.

Whether these functions remain approximately constant or not is controlled by an adiabaticity condition (see~\cite{Barcelo:2010pj,Barcelo:2010xk}) which, in its simplest form, can be written as~$|\dot{\kappa}|/\kappa^2 \ll 1$ (either for~$\kappa_U$ or~$\kappa_V$) for a period of time~$\Delta \tau \gtrsim 1/|\kappa|$, where the dot denotes derivative with respect to~$\tau$. Physically, this condition ensures that the function does not significantly change during the time needed to detect a particle of the energy characteristic of the planckian spectrum. When this condition holds, one can replace this approximately constant value of~$\kappa$ in its own definition~(\ref{kappa}) and integrate the equation in the interval~$\Delta \tau$, obtaining~$U(\tau) \approx U_0 - A \rme^{-\kappa_{_U} \tau}$ (and equivalently for~$V(\tau)$). This exponential relation between the affine parameters associated with the observer and with the vacuum state is the necessary condition for the observer to detect a planckian spectrum of particles along the given interval.

\section{Hawking versus Unruh effects}\label{sec:H-vs-U}

As already mentioned in the introduction, in this work we shall propose a clean separation between the Hawking and Unruh effects (under the simplifying assumptions of the analysis). This separation amounts to distinguishing between \emph{radiation action} as the Hawking effect, and \emph{radiation back-reaction} as the Unruh effect (an interpretation which we already advanced in~\cite{Barbado:2015zga}---see also~\cite{Barbado2016}). In the first case, it is the radiation emitted by the black hole and already present in the field which \emph{acts} upon the detector; while in the second, it is the detector that perturbs the field and, by \emph{back-reaction,} modifies its trajectory at the same time that it becomes excited.

A particle detector is sensitive to both the quantum state of the field and the detector trajectory. In general situations, one might not be able to distinguish the origin (radiation action or radiation back-reaction) of a particular excitation of the detector. However, at least when neglecting back-scattering in the spherically symmetric and asymptotically flat geometries of our interest, this distinction is possible. If a detector in a radial trajectory becomes excited, the energy has to be supplied by either: i) the energy content in the quantum field, ii) the energy supplied by the detector's rockets, or iii) the gravitational potential energy of the detector in the curved geometry. The first source is the only source of radiation action, while the other two are clearly sources of radiation back-reaction. In the presence of an asymptotically flat region, we may consider a static detector there (which is also arbitrarily close to inertial) as a reference detector. This detector has no rockets and does not exploit its gravitational energy. Thus, its excitations must be due exclusively to particles already present in the quantum field, i.e.\ to radiation action or Hawking effect. Comparison with respect to this reference detector will then allow us to clearly identify the character (action or back-reaction, Hawking or Unruh) of the different components in any scenario.

Specifically, we propose to associate the Unruh effect with the acceleration of the detector with respect to the asymptotic region, and not with respect to the local free-fall reference as is usually understood. To further understand the reasons, it is interesting to compare the Unruh effect with radiation reaction effects in classical electrodynamics in black-hole background spacetimes. A classical point charge at rest at a fixed radial position in a Schwarzschild black hole does not radiate and so does not experience any radiation-reaction effect~\cite{Boulware1980} (the charge does experience a dipolar force due to a polarization of the horizon though~\cite{SmithWill1980}, but this is an entirely different effect). This is so because it does not change the structure of the electromagnetic field at infinity. On the contrary, a charge moving along a radial geodesic (that is, in the background free fall) does experience radiation reaction forces~\cite{DeWitt1964}. Any acceleration with respect to the asymptotic region is a cause of radiation and radiation-reaction, independently of whether it is caused by gravity or by any other force field. This same idea is here transferred to our interpretation of the Unruh effect. To be more concrete let us describe the separation formula between Hawking and Unruh effects.

\subsection{Separation formula}

For concreteness, let us consider the radial sector of a Schwarzschild geometry (although the calculations can easily be extended to other forms of spherically symmetric black holes). In~\cite{Barbado:2012pt}, under the assumption of ignoring backscattering (thus reducing the problem to the conformal $(1+1)$-dimensional quantum field theory described here), a formula was derived to express the effective temperature function $\kappa_U (\tau)$ in the outgoing radiation sector perceived by an observer located at a radial position~$r$ with a local velocity~$\vl$ with respect to the black hole, and a proper acceleration~$\apr$, namely:
\begin{equation}
\kappa_U (\tau) = \sqrt{\frac{1-\vl}{1+\vl}} \frac{1}{\sqrt{1-\frac{2M}{r}}} \left(\ks_U (\ub (\tau))-\frac{M}{r^2}\right) + \apr,
\label{local_kappa}
\end{equation}
where~$M$ is the mass of the black hole, while $\vl$ and $\apr$ can be written in terms of the radial coordinate $r$ and its $\tau$-derivatives:
\begin{equation}
\vl = \frac{\dot{r}}{\sqrt{1-2M/r+\dot{r}^2}}, \quad \apr = \frac{\ddot{r}+M/r^2}{\sqrt{1-2M/r+\dot{r}^2}}.
\label{def_v_a}
\end{equation}
Finally, the \emph{state effective temperature}~$\ks_U (\ub)$ is
\begin{equation}
\ks_U (\ub) := -\left. \frac{\rmd^2 U}{\rmd \ub^2} \middle/ \frac{\rmd U}{\rmd \ub} \right.,
\label{kappa_st}
\end{equation}
where~$\ub := t - r^*$ is now the outgoing Eddington-Finkelstein null coordinates ---the ingoing one being~$\vb := t + r^*$---, with~$r^* = r + 2M \log[r/(2M)-1]$ the \emph{tortoise coordinate.} Thus, $\ks_U (\ub)$ only depends on the vacuum state of the field, and not on the observer's radial position or motion. Notice that the reference observer in the asymptotic region described earlier naturally couples to the modes~$\{\rme^{\rmi \omega \ub},\rme^{\rmi \omega \vb}\}$, so that, \emph{for him}, \mbox{$\rmd \ub = \rmd \vb = \rmd \tau$}. Thus, $\ks_U (\ub)$ is the effective temperature function which the reference observer perceives when crossing the outgoing ray~$\ub$. This ray is crossed by our generic observer at his proper time~$\tau$, given~$\ub (\tau)$. This fact suggests the following separation of formula~(\ref{local_kappa}) into Hawking and Unruh parts:
\begin{align}
\kappa_U (\tau) = & \kappa^{\rm Haw}_U (\tau) + \kappa^{\rm Unr}_U (\tau), \label{kappa_u_separation}
\\
\kappa^{\rm Haw}_U (\tau) := & \sqrt{\frac{1-\vl}{1+\vl}} \frac{1}{\sqrt{1-\frac{2M}{r}}} \ks_U (\ub(\tau)), \label{kappa_u_hawking}
\\
\kappa^{\rm Unr}_U (\tau) := & - \sqrt{\frac{1-\vl}{1+\vl}} \frac{1}{\sqrt{1-\frac{2M}{r}}}~\frac{M}{r^2} + \apr. \label{kappa_u_unruh}
\end{align}
Indeed, the Hawking radiation part corresponds to the radiation detected in the asymptotic region, which only depends on the state of the field, adequately shifted by a Doppler factor and a gravitational blueshift corresponding to the observer's velocity and position, respectively. The Unruh part, on the other hand, depends only on the state of motion with respect to the black hole---or, equivalently: with respect to the asymptotic region---, but is totally independent of the quantum state of the field. Notice that, depending on the sign of each contribution, Hawking and Unruh effect can contribute oppositely to the final temperature perceived~(\ref{kappa_u_separation}), leading to a lower temperature than that of one of the effects separately, or even to its total cancellation. We interpret this as a destructive interference between the effects, and we will describe a particularly important example of it later on.

The separation in~(\ref{kappa_u_separation})--(\ref{kappa_u_unruh}) could also be written [by using the chain rule in~(\ref{kappa})] as
\begin{align}
\kappa^{\rm Haw}_U (\tau) = & \frac{\rmd \ub}{\rmd \tau} \ks_U (\ub(\tau)), \label{kappa_u_hawking_alt}
\\
\kappa^{\rm Unr}_U (\tau) = & -\left. \frac{\rmd^2 \ub}{\rmd \tau^2} \middle/ \frac{\rmd \ub}{\rmd \tau} \right.. \label{kappa_u_unruh_alt}
\end{align}
Equation~(\ref{kappa_u_unruh_alt}) makes explicit that the Unruh effect, in our interpretation, is due to the acceleration \emph{with respect to the asymptotic region,} and not with respect to the local free-falling frame. Actually, the previous formulas~(\ref{kappa_u_hawking_alt}) and (\ref{kappa_u_unruh_alt}) are general formulas for a spacetime with an asymptotically flat region in which the affine outgoing null coordinate is~$\ub$. Only accelerations with respect to the asymptotic region, which can alter the behavior of the field at infinity, can involve exchanges of energy with the asymptotic region, and thus radiation-reaction effects. This acceleration is determined by the rate of change of the frequency shift~$\rmd \ub / \rmd \tau$ between the observer and the asymptotic region, and it can be due to the proper acceleration of the observer, but also due to the gravitational attraction of the black hole, as Eq.~(\ref{kappa_u_unruh}) clearly shows.

When the field is in the Boulware state (associated with the Eddington-Finkelstein null coordinates, i.e. $U = \ub$), then $\ks_U(\ub) = 0$ (see Eq.~\eqref{kappa_st}). In that case,~\eqref{kappa_u_unruh_alt} coincides with the definition of the total $\kappa_U(\tau)$. Thus the black hole emits no radiation and any radiation detected must be due purely to the Unruh effect.

Finally, we can obtain equivalent expressions for the ingoing radiation sector:
\begin{align}
\kappa_V (\tau) = & \kappa^{\rm Haw}_V (\tau) + \kappa^{\rm Unr}_V (\tau), \label{kappa_v_separation}
\\
\kappa^{\rm Haw}_V (\tau) := & \sqrt{\frac{1+\vl}{1-\vl}} \frac{1}{\sqrt{1-\frac{2M}{r}}} \ks_V (\vb(\tau)), \label{kappa_v_hawking}
\\
\kappa^{\rm Unr}_V (\tau) := & \sqrt{\frac{1+\vl}{1-\vl}} \frac{1}{\sqrt{1-\frac{2M}{r}}}~\frac{M}{r^2} - \apr. \label{kappa_v_unruh}
\end{align}
Here, $\ks_V (\vb)$ is defined in a way analogous to~$\ks_U (\ub)$ in~(\ref{kappa_st}). The changes of signs in~\eqref{kappa_v_hawking} and~\eqref{kappa_v_unruh} as compared with~\eqref{kappa_u_hawking} and~\eqref{kappa_u_unruh} can be easily explained by noticing that the change~$\ub \to \vb$ is mathematically equivalent to~$r^* \to - r^*$, which itself is equivalent to the changes~$r \to -r$ and~$M \to -M$. Looking also at the expressions in~(\ref{def_v_a}), one can fully explain the changes of signs obtained.

Note that this separation between Hawking and Unruh components of the radiation will not affect the calculation of the temperature measured by a detector. Still, our proposal is not merely a matter of interpretational consistency. It becomes relevant when one wants to calculate effects beyond the temperature measured by a detector (in particular, effects related to radiation action and back-reaction). In such case, different interpretations could suggest different predictions and misguide a calculation. In the following we will illustrate this fact with two examples of observers, and discuss its possible implications for black hole physics.

\section{Observers outside a black hole}\label{sec:examples}

\subsection{Static observers}

In this and the next section, we will use the effective temperature function to analyze the perception of static and free-falling observers outside a Schwarzschild black hole (see also~\cite{Barbado:2011dx}). The quantum field will be in the Unruh vacuum state, which correctly describes the aftermath of a black-hole collapse~\cite{Unruh:1976db}. Since~$\ks_U = 1/(4M)$ and~$\ks_V = 0$ in this state, there is no ingoing radiation in the asymptotic region, but only a flux of outgoing thermal radiation at the Hawking temperature
\begin{equation}
T_{\rm H} = \frac{|\ks_U|}{2\pi} = \frac{1}{8\pi M}.
\label{temp_hawking}
\end{equation}

An observer at a constant radial position~$\rs$ outside a black hole has no acceleration with respect to the asymptotic region. We therefore expect no Unruh effect at all. Indeed, for the observer to be static it requires~$\vl = 0$ and~$\apr = M/\Big(\rs^2 \sqrt{1-2M/\rs}\Big)$, so that~$\apr$ compensates for the gravitational pull of the black hole. One then sees from~(\ref{kappa_u_separation})--(\ref{kappa_u_unruh}) and~(\ref{kappa_v_separation})--(\ref{kappa_v_unruh}) that
\begin{equation}
\kappa^{\rm Unr}_U = \kappa^{\rm Unr}_V = 0,
\end{equation}
while
\begin{equation}
\kappa_U = \kappa^{\rm Haw}_U = \frac{1}{\sqrt{1-\frac{2M}{\rs}}} \frac{1}{4M}, \qquad \kappa_V = 0.
\label{kappa_static}
\end{equation}
The observer perceives only outgoing Hawking radiation with a gravitational blueshift dependent on his radial position. This radiation will produce a buoyancy effect and could, under certain circumstances, even be sufficient to maintain an observer at a fixed radial position~\cite{Barbado:2015zga}. The ingoing sector, on the other hand, is completely void of radiation.

It is important to recall that the fact of experiencing this huge particle perception when close to the horizon does not imply that Hawking particles originate at the horizon (see e.g. the recent discussion in~\cite{Giddings2015}). For instance, the renormalized stress-energy tensor in the Unruh state is negligible at the horizon of an astrophysical black hole, so close to the horizon there is no real energy density associated to outgoing radiation. It is only the perception that becomes large at the horizon. For a clear distinction between perception and objective quantities we refer the reader to a previous paper of the authors~\cite{Barbado2016}. Another way to look at this issue is by separating particle production from vacuum polarization effects (see e.g. the recent paper~\cite{Firouzjaee2015}).

\subsection{Free-falling observers}

\begin{figure}[h]
\centering
\includegraphics[width=8cm]{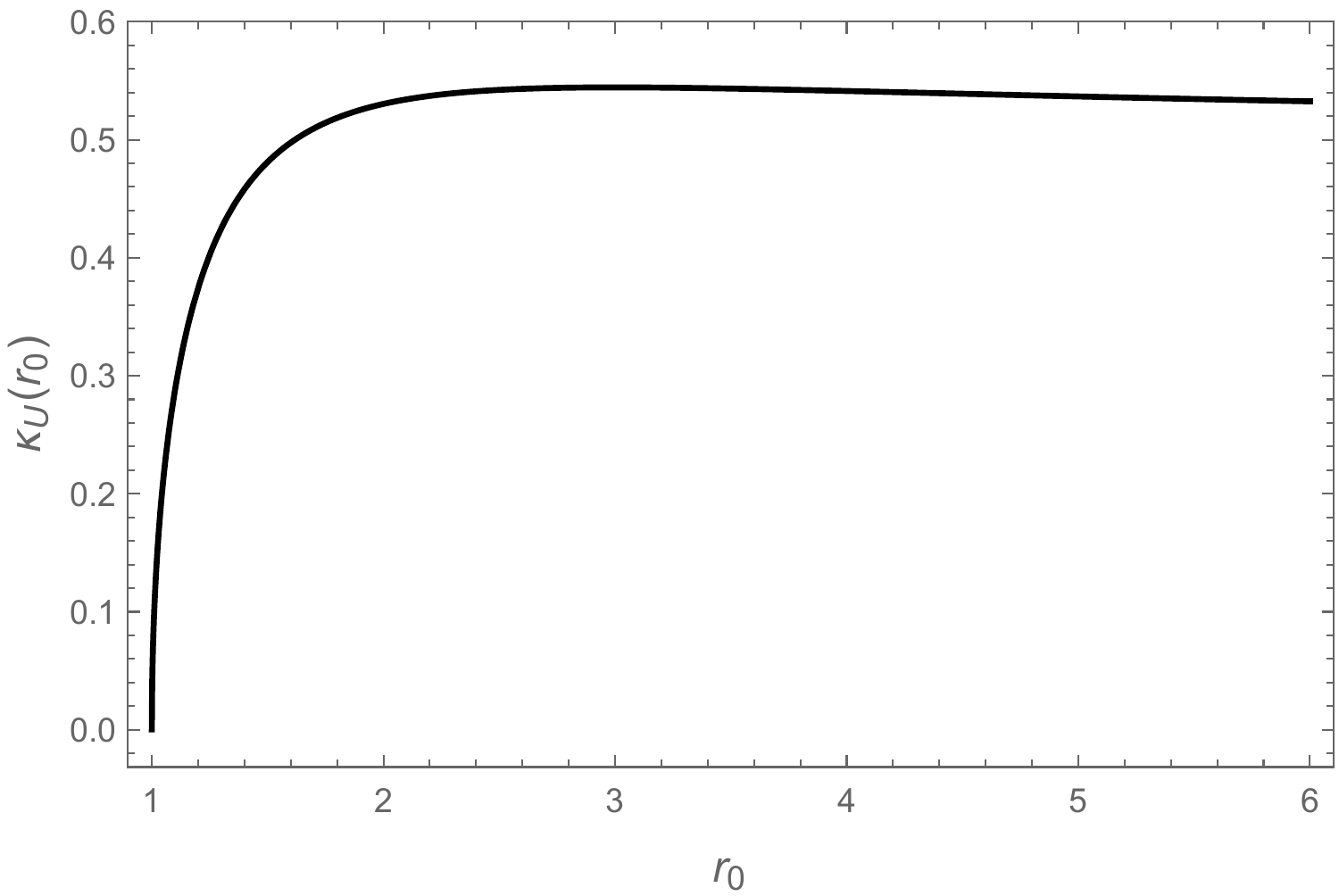}
\caption{Effective temperature function $\kappa_U$ as a function of the radius~$r$ for free-falling and instantaneously static observers in the Unruh vacuum state. We use $2M=1$~units.}
\label{fig_unruh}
\end{figure}

For a free-falling observer ($\apr = 0$) with zero velocity ($\vl = 0$); that is, ``just released'' into free-fall at some radius~$r = r_0$, we have that, in the outgoing sector,
\begin{equation}
\kappa^{\rm Haw}_U = \frac{1}{\sqrt{1-\frac{2M}{r_0}}} \frac{1}{4M},\qquad\ \kappa^{\rm Unr}_U = - \frac{1}{\sqrt{1-\frac{2M}{r_0}}}~\frac{M}{r_0^2}.
\end{equation}
The Hawking and Unruh components both increase (in absolute terms) as the release point approaches the black hole, and diverge at the horizon. However, the opposite signs mean that they interfere destructively, leading to a small net perception of
\begin{equation}
\kappa_U = \kappa^{\rm Haw}_U + \kappa^{\rm Unr}_U = \frac{1}{\sqrt{1-\frac{2M}{r_0}}} \left( \frac{1}{4M} - \frac{M}{r_0^2} \right),
\end{equation}
which vanishes as the release point reaches the horizon (see~Fig.~\ref{fig_unruh}).
This result requires not only that the observer be in free fall, but also that he be ``instantaneously static'' ($\vl = 0$). However, this is not the case when we consider the complete trajectory for an observer released to free fall from a finite distance~$r_0 > 2M$ beyond the horizon: he will immediately acquire a non-zero velocity~$\vl$, which will tend to~$\vl \to -1$ when crossing the horizon [see~(\ref{def_v_a})]. This yields a diverging Doppler factor $\sqrt{\frac{1-\vl}{1+\vl}}$, which reinforces the divergence of both the Hawking and the Unruh components [(\ref{kappa_u_hawking}) and~(\ref{kappa_u_unruh})] at the horizon, but again with opposite signs. The resulting effective temperature at the horizon crossing is in this case non-zero, but finite and given by \cite{Barbado:2011dx}
\begin{equation}
\kappa^{\rm hor}_U = \sqrt{1-\frac{2M}{r_0}} \frac{1}{M}.
\label{kappa_hor_u}
\end{equation}
This effective temperature can be up to $4 T_{\rm H}$ for observers released arbitrarily far from the horizon ---which is still a tiny radiation---, and tends to zero for observers released close to the horizon, in agreement with the well-known vanishing result obtained previously.

Let us now look at the ingoing sector for these free-falling observers [see (\ref{kappa_v_separation})--(\ref{kappa_v_unruh})]. A free-falling observer with an instantaneous velocity $\vl = 0$ sees
\begin{eqnarray}
\kappa_V = \kappa^{\rm Unr}_V = \frac{1}{\sqrt{1-\frac{2M}{r_0}}}~\frac{M}{r_0^2}.
\end{eqnarray}
There is no Hawking effect in the ingoing sector for the Unruh vacuum: $\kappa^{\rm Haw}_V=0$. However, the observer does experience a strong Unruh effect in the ingoing sector, actually equal in absolute value to that in the outgoing sector but with opposite sign, which diverges as he approaches the horizon. In the Unruh vacuum state, the ingoing sector is perceived as vacuum by observers at fixed radial positions, just as in the Boulware vacuum state. It therefore makes sense that a free-falling detector would indeed detect particles in this sector.

\begin{figure}[h]
\centering
\includegraphics[width=8cm]{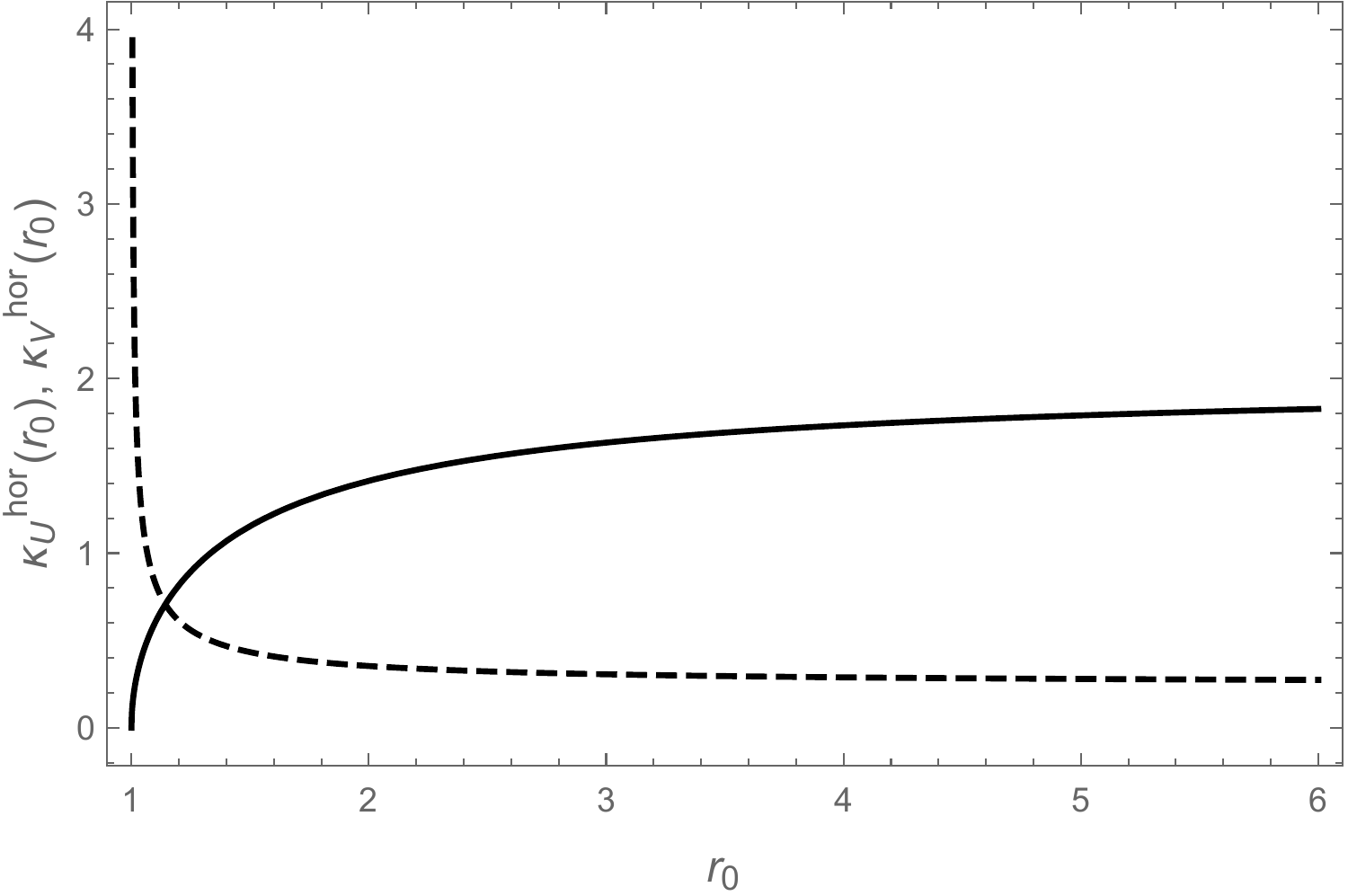}
\caption{Horizon-crossing values of the effective temperature functions $\kappa^{\rm hor}_U$ (solid line) and~$\kappa^{\rm hor}_V$ (dashed line) as a function of the releasing radius~$r_0$ for free-falling observers in the Unruh vacuum state. We use $2M=1$~units.} 
\label{fig_horizons}
\end{figure}

However, when one considers observers released from a finite distance beyond the horizon ($r_0 > 2M$), the ingoing Doppler factor $\sqrt{\frac{1+\vl}{1-\vl}}$ goes to zero when crossing the horizon ($\vl \to -1$) and tends to compensate the diverging result above. The resulting effective temperature is indeed again finite at the horizon crossing, and given by
\begin{equation}
\kappa^{\rm hor}_V = \frac{1}{\sqrt{1-\frac{2M}{r_0}}} \frac{1}{8M}.
\label{kappa_hor_v}
\end{equation}
It can be up to $T_{\rm H}/2$ for observers released far from the horizon, and diverges for observers released close to the horizon, in agreement with the previous result. In~Fig.~\ref{fig_horizons} we plot the horizon-crossing values~(\ref{kappa_hor_u}) and~(\ref{kappa_hor_v}) as a function of the releasing radius~$r_0$.

The interpretation advocated in this article makes clear that Unruh and Hawking effects are quite distinct. In~\cite{Singleton:2011vh} and more recently in~\cite{Singleton2016} it is argued that the principle of equivalence is violated in quantum radiating black holes. They compare the particle perception of an observer at a fix radius in a black hole spacetime in the Unruh state and an uniformly accelerating observer in a Minkowski spacetime in the Minkowski vacuum, considering that both situations are parallel and should not be distinguishable if the equivalence principle were satisfied. However, within our interpretation it is clear that both situations should not be compared from the equivalence principle perspective. The first situation corresponds to a pure Hawking effect, zero in the ingoing sector and non-zero in the outgoing sector. The second corresponds to a pure Unruh effect, non-zero in both the ingoing and outgoing sectors. But moreover, in the first situation some radiation is being thrown from the black hole towards the observer, which is accelerating with respect to the free-fall reference. In the second situation the accelerated observer is not subjected to any external source of radiation. Thus, both situations cannot be compared because represent rather different physical situations (a criticism along the same line was already offered in~\cite{Crispino:2012zz}).

\section{The difficulty of crossing the horizon at low velocities}\label{sec:difficult}

In~\cite{Barbado:2011dx} we stressed that one has to be careful with blindly using the usual statement ``the Unruh state is seen as vacuum by free-falling observers at the horizon''.

Analyzing the outgoing sector alone, we showed that the common belief that a free-falling observer near the horizon experiences a vacuum state, only holds strictly true when he is released into free fall extremely close to the horizon, but not for larger initial distances $r_0$. In the case of static observers, it is also in that limit, when the position~$\rs$ of the observer is close to the horizon, when the value of~$\kappa_U$ in~(\ref{kappa_static}) really approaches the proper acceleration~$\apr$ of the observer. This explains why the ``traditional'' interpretation that ``Hawking radiation reduces to the Unruh effect near the horizon'' gives the correct result for the temperature measured near the horizon. However, for general release positions of free-falling observers (not necessarily close to the horizon), what is nonetheless still true, is that even though a detector will perceive something in the outgoing sector, this perception will have the same order of magnitude as the Hawking flux at infinity and so will be extremely tiny for astrophysical black holes. In the outgoing sector, no high-energy phenomena will be experienced by a free-falling detector, in agreement with the standard paradigm.

Apart from the well-known results of the outgoing sector, in this work we have also analyzed the ingoing sector. Again, free-falling observers released from large initial radial positions~$r_0$ perceive ingoing particles with intensities of the order of the asymptotic Hawking radiation. However, when the detector is released from positions already close to the horizon (equivalent to saying that the horizon will be crossed at low velocities), while the perception in the outgoing sector becomes even more negligible, the perception in the ingoing sector becomes enormous. In our interpretation this is due to the Unruh effect ---now meaning acceleration with respect to the asymptotic region--- and thus it should work against the gravitational acceleration through back-reaction of the emitted Unruh radiation, forcing the trajectory of the detector to slow down, and hence to strongly deviate from free fall.   

The process is the following: For initially free-falling observers with low velocity nearby the horizon, a huge Unruh effect appears in the ingoing sector which tends to brake the free-falling trajectory. But, once the observer, due to this fact, becomes \emph{not-so-free-falling,} and thus the Unruh effect interacting destructively with the Hawking radiation in the outgoing sector diminishes, he must start perceiving the Hawking radiation emitted by the black hole with a huge gravitational blueshift. And this radiation would produce a radiation pressure force leading the trajectory to deviate even further from free fall. This buoyant force can even create the possibility of an observer remaining in a stationary radial position outside the black hole horizon, a buoyant point. The huge ingoing Unruh effect has disappeared and been substituted by a huge outgoing Hawking effect. Another way to describe this is that, in terms of perception~\cite{Barbado2016}, a huge ingoing flow of negative energy becomes a huge outgoing flux of positive energy. Using this different approach, the authors in~\cite{Eune:2014eka} also identified potential difficulties when crossing the horizon.

All in all, our analysis points out that the standard picture in which crossing the horizon can be done without encountering high-energy effects, is true but only under the qualification that the horizon cannot be approached at low velocities. If something happens to a free-falling detector before crossing the horizon so that its velocity is strongly decreased (e.g., its rockets are switched on for a moment), these described effects could take over and prevent the detector from crossing the horizon.

The situation is reminiscent of what happens with the very collapse of matter to form a black hole in the first place. In~\cite{Barcelo2008} it was highlighted that for strong quantum effects not to affect the classical collapse process before horizon formation, it is necessary that the collapsing matter crosses its gravitational radius sufficiently fast. This is indeed what is supposed to happen in the standard collapse processes we know about. However, it is important to keep this typically hidden assumption in mind when analyzing alternative models of collapse~(see e.g. the discussions in~\cite{Barcelo2015,Barcelo2016}). This same attitude is the one we maintain with the present work.

\section{Summary}\label{sec:summary}

In this paper we have proposed a consistent interpretation of what constitutes Hawking radiation and what constitutes an Unruh effect. Under the simplifying assumptions of having an asymptotically flat region, spherically symmetric configurations, and negligible back-scattering, we have been able to provide an analytic formula separating these two effects. In this interpretation the Unruh effect is associated with the acceleration of a detector with respect to the asymptotic region and not with respect to the local free-fall reference frame. We associate the Hawking and Unruh effects with radiation action and radiation back-reaction effects, respectively.

Armed with this interpretation, we have calculated the Hawking and Unruh effects in standard radiating black holes. We point out that to say that a ``detector in free fall can cross the horizon without experiencing any high-energy effect'' needs a further qualification: that the crossing is not attempted at low velocities. In the case of low velocities, we argue that a strong Unruh effect will appear in the ingoing sector which through a dampening process can drastically affect the detector's trajectory and could even prevent the detector from crossing the horizon altogether.

\section*{Acknowledgments}

Financial support was provided by the Spanish MINECO through the projects FIS2014-54800-C2-1, FIS2014-54800-C2-2 (with FEDER contribution), and by the Junta de Andaluc\'{\i}a through the project FQM219.

\end{document}